\documentstyle{aipproc}
\input epsfig.sty
\def\gr{$\gamma$-ray }

\begin{document}
\title{Maximum-Entropy analysis of EGRET data}
\author{M. Pohl$^{1,2}$ for the EGRET collaboration, and A.W. Strong$^1$}
\address{1 MPI f\"ur Extraterrestrische Physik, Postfach 1603, 85740 Garching,
Germany; \\
2 Danish Space Research Institute, Juliane Maries Vej 30, 2100 K\o benhavn \O,
Denmark}

\maketitle
\begin{abstract}
EGRET data are usually analysed on the basis of the Maximum-Likelihood
method \cite{ma96} in a search for point sources in excess
to a model for the background radiation (e.g. \cite{hu97}).
This method depends strongly on the quality of
the background model, and thus may have high systematic uncertainties
in region of strong and uncertain background like the
Galactic Center region.

Here we show images of such regions obtained by the quantified 
Maximum-Entropy method. We also discuss a possible further use of
MEM in the analysis of problematic regions of the sky.
\end{abstract}
\section{The method}
Scientific data analysis means using our data to make inferences about
various hypotheses. However, what the data give us is the likelihood
of getting our specific data set as a function of the hypothesis.
To
reverse the conditioning Bayes' theorem can be used which invokes
a prior probability distribution for the hypotheses as additional input.
The prior probability distribution
can be fairly simple, for example a scaling of a
known behaviour in a system to its subsystem.

The consequence of this is that the "best" hypothesis can be derived
by maximising the entropy (for an extensive introduction see \cite{gs84}) 
\begin{equation}
S(h)=\sum_{i=1}^L (h_i -m_i -h_i \log({{h_i}\over {m_i}}))
\end{equation}
under the constraint of a likelihood argument.
Here $h_i$ is the hypothesis in pixel $i$ and $m_i$ is our prior
expectation. 

An implementation of this method is given by the 
MEMSYS5-package, which further allows us to derive a posterior probability
bubble and thus the uncertainty of the "best" hypothesis.

\section{The influence of the prior expectation}
From the definition it is clear that the prior expectation of the
analyst has a strong influence on the result. This implies also that
without information on what kind of prior was used the results of a
MEM run have to be taken with care.

\begin{figure}
\centerline{\epsfig{file=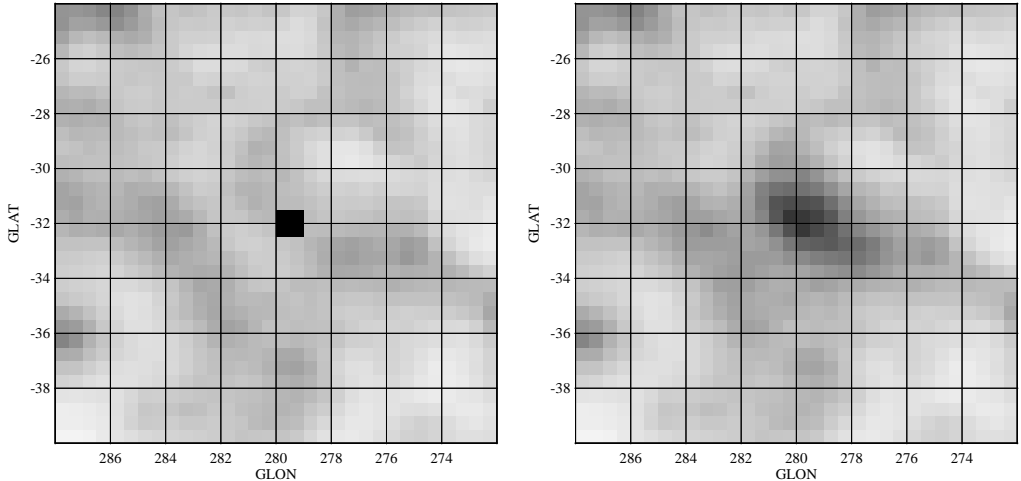}}
\caption{The sky region around LMC as seen by EGRET. The grey scale
is linear in integrated intensity above 100 MeV. The image on the
right side is the result of a Maximum-Entropy deconvolution with
a model of the galactic diffuse emission as prior expectation,
whereas on the left side we see the image resulting when a point
source at the position of LMC is added to the prior expectation.
Though the right image seems to indicate a substantial spatial extent 
of the LMC emission, the data seem to be compatible with LMC being
1$^\circ$ in size or less.}
\label{lmc}
\end{figure}

It is often argued that MEM is advantageous for the analysis of
diffuse emission or extended structure. In Fig.\ref{lmc} we show
the region around the Large Magellanic Cloud (LMC) twice. One
image is derived without having LMC included into the prior
expectation, and the second is the result of adding
a point source to the prior expectation based the flux and best
position of LMC \cite{th95}. Though the prior distribution is
different only in an area of 1 square degree, the final image is
influenced over roughly 20 square degrees. This implies that
a) the appearance of LMC depends strongly on the prior expectation, and
b) any source of unknown extent, which is not accounted for in the
prior distribution, may result in a fuzz of a few degrees extent.
So with MEM we do find sources, but it is hard to distinguish point
sources from structure like that in the sky distribution of galactic
diffuse emission. 

\begin{figure}
\centerline{\epsfig{file=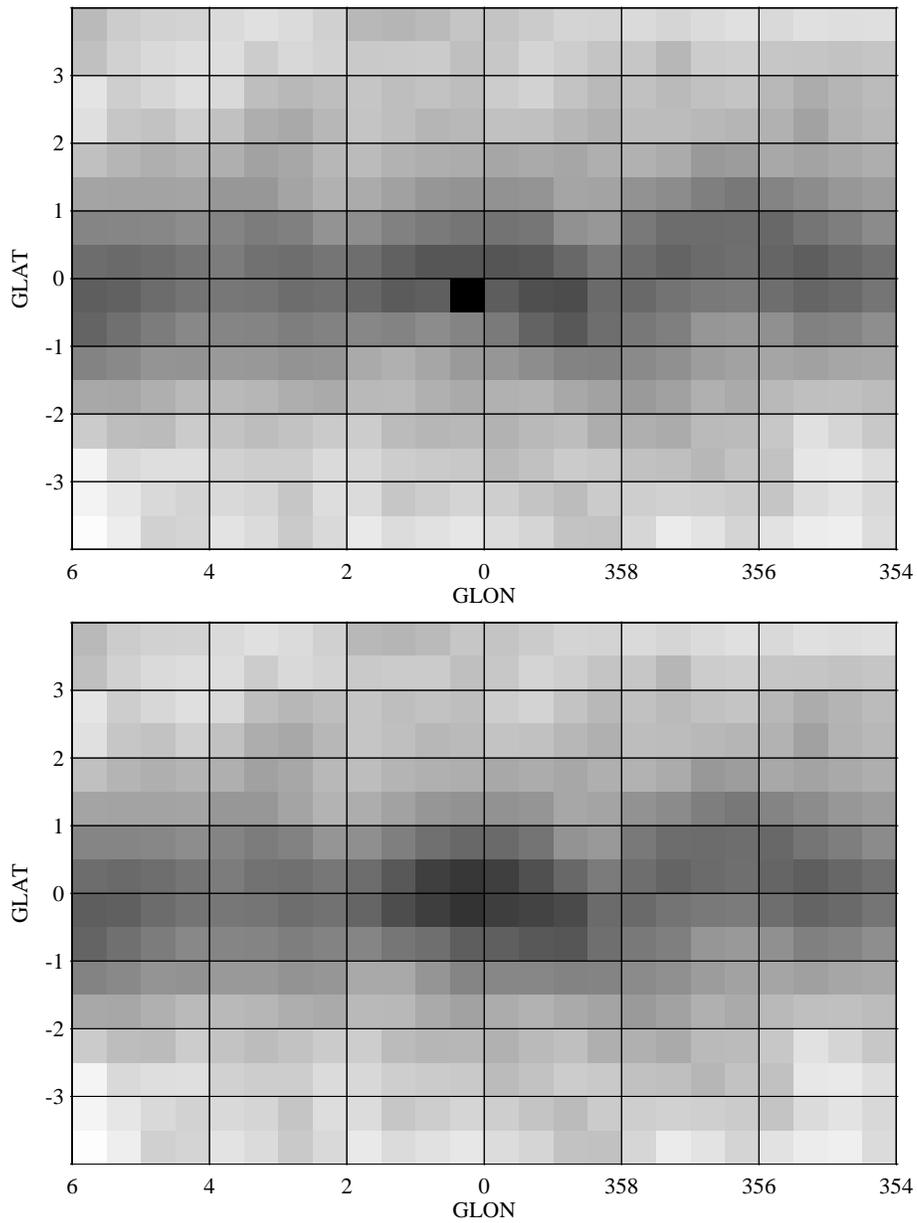}}
\caption{The galactic center region at energies above 300 MeV displayed
in a linear grey scale.
The upper image is derived with a model of the galactic diffuse emission
and a galactic center point source in the prior distribution,
whereas for the
lower image only the model of diffuse emission has been used to
construct the prior distribution. As in Fig.1 the appearance of
the Galactic Center source depends strongly on our expectation.}
\label{gc}
\end{figure}
A similar effect is seen in crowded regions like the Galactic Center,
as shown in Fig.\ref{gc}. Here we have restricted ourselves to
\gr energies above 300 MeV so that the smaller extent of the
point-spread function can be used to advantage. The second EGRET
catalogue \cite{th95} lists a highly significant source right
at the position of the Galactic Center, but notes potential
problems of the likelihood analysis in this region. In the MEM analysis
we again see a strong influence of the prior model on the resulting
image, however on smaller scale. This is due both to our restriction 
to higher energies and to the better statistic than in case of LMC.

When we now artificially increase the flux of the galactic center
source in the data, but still use the catalogue flux in our prior
distribution, then the additional intensity will be spatially distributed
like in the case of no point source in the prior, i.e. the lower panel
of Fig.\ref{gc}. 

\section{Discussion}
We have shown that on small scales the MEM images tend to be influenced
strongly by the prior expectation. Thus point sources can not be easily 
distinguished from structure in the diffuse emission. Also, in single
viewing periods the prior distribution is likely to be dominant of the
likelihood statistic and thus the final image. Thus MEM can not be used
in a simple way to deduce the light curve of point sources
in regions of high confusion level. 
However, what we can
do is to use MEM to improve the background model for the likelihood
analysis of known point sources in crowded regions of the \gr sky.

Despite our concerns about the limited reliabity of individual
structure in the MEM output, it is still true that MEM produces 
images which are highly reliable on larger scales. As an example
we show an allsky image in Fig.3. Please note that the
grey scale here is logarithmic and spans more than two orders of
magnitude in integrated intensity above 100 MeV. The point sources
stick
out very clearly. None of the point sources is in the prior 
distribution, so that
all of them have an extent of a few degrees in the image. 

\begin{figure}
\centerline{\epsfig{file=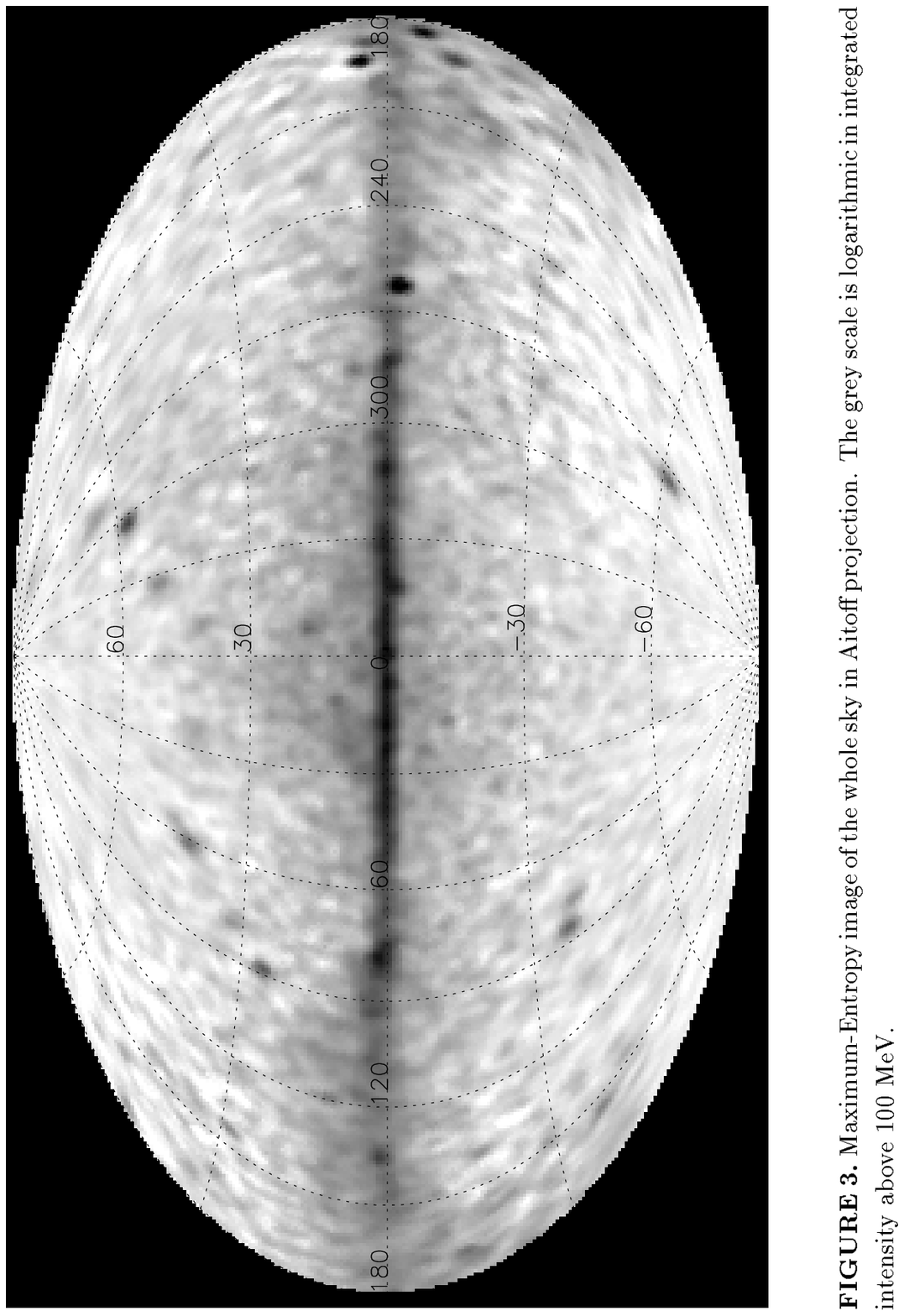}}
\end{figure}
\vskip0.5truecm
\noindent
The EGRET Team gratefully acknowledges support from the following:
Bundesministerium f\"{u}r Bildung, Wissenschaft, Forschung und Technologie
(BMBF), Grant 50 QV 9095 (MPE); NASA Cooperative Agreement NCC 5-93 (HSC); 
NASA Cooperative Agreement NCC 5-95 (SU); and NASA Contract NAS5-96051
(NGC).

\end{document}